\documentstyle[11pt]{article}

\begin{document}
\Large \bf

\begin{center}
Quantum deflation of classical extended objects
\end{center}

\begin{center}
K.Sveshnikov\footnote{\large E-mail: costa@bog.msu.su}
\end{center}

\large \it
\begin{center}
Department of Physics and Institute of Theoretical Microphysics,
\end{center}
\begin{center}
Moscow State University, Moscow 119899, Russia
\end{center}

\vskip 0.3 true cm

\leftskip 1 true cm
\rightskip 1 true cm
\large \rm
\baselineskip 10 pt
It is shown, that extended particle-like
objects should infinitely long collapse into some discontinuous
configurations of the same topology, but vanishing mass.
 Analytic results concerning the general properties and
asymptotic rates of such a process are given for
1+1-dimensional soliton models.

\vskip 0.3 true cm

\begin{center}
{hep-th/9410160}
\end{center}

\vskip 0.3 true cm

\leftskip 0 true cm
\rightskip 0 true cm
\baselineskip 11pt
\Large \rm

\def\prop{\hbox{\simv \char 47}}
\def\lagr{\hbox{$\cal L$}}
\def\ham{\hbox{$\cal H$}}
\def\C{\hbox{Const}}
\def\dalam{\hbox
{\vrule\vbox{\hrule\hbox to 1ex{ \hfill}\kern 1 ex\hrule}\vrule}}
\def\1/2{\hbox{$ {1 \over 2}$ }}
\def\i/h{{i \over \h}}
\def\f{\varphi} \def\F{\Phi}
\def\e{\varepsilon} \def\E{\hbox{$\cal E $}}
\def\<{\langle}
\def\>{\rangle}
\def\th{\tanh}
\def\ch{\cosh}
\def\sh{\sinh}
\def\sign{\hbox{sign}}
\def\h {\hbar}
\def\a{\alpha}
\def\b{\beta}
\def\g{\gamma}  \def\G{\Gamma}
\def\d{\delta}  \def\D{\Delta}
\def\l{\lambda}   \def\L{\Lambda}
\def\s{\sigma}
\def\r{\varrho}
\def\x{\xi}
\def\c{\chi}
\def\m{\mu}
\def\n{\nu}
\def\t{\tau}
\def\p{\pi}
\def\k{\kappa}
\def\z{\zeta}
\def\({\left(}
\def\[{\left[}
\def\){\right)}
\def\]{\right]}
\def\pd{\partial}
\def\w{\omega}

At present, classical particle-like solutions (henceforth solitons) play a
significant role in the QFT, since they are considered as reliable
models of quantum extended objects, e.g. existing baryons [1-3].
It is widely believed, that if the classical soliton  is stable due to
either topological or dynamical reasons, then upon quantization by
existing techniques its quantum descendant should most likely reveal the
same stability properties [1-3,6-9,11-15].
The purpose of this paper is to present arguments, that in the QFT, under
rather general circumstances, such a correspondence cannot take place.
Namely, we'll describe a peculiar quantum effect, that makes
possible a special kind of collapse of such extended objects into certain
almost discontinuous configurations of the same topology, but vanishing
mass.

Such configurations appear
quite naturally within the explicitly lo\-re\-ntz-co\-va\-ri\-ant treatment of
the soliton Heisenberg field by means of Newton-Wigner position
variables [4]. The latter emerge as the direct result of
summing up the pertinent terms in the perturbative solution of eqs. of
motion starting from the static classical solution [5] and
serve as the covariant collective coordinates of the soliton, which
restore the broken Lorentz symmetry and provide the cancellation of
corresponding zero modes [6-9].
These covariant coordinates give rise to an effective dynamics,  where all
the differential operators of the initial theory are replaced by
finite differences with the step, proportional   to
the (effective) Planck constant of the theory $\h$ [4,10].  In turn, such
effective theory yields a series of c-number copies of classical solution,
labeled by an integer $N$, which are quite analogous to and reveal the
same topological properties as the initial soliton, whereas their masses
are much less than the classical one and vanish for $\h/N \to 0$ [4].
 Within such a picture
 the soliton collapse appears quite naturally as a transition from the
classical soliton state into its Nth copy with the same topological
charge.  The peculiarity of this process is, that the
dynamics of the collapse can be successfully understood in the
quasiclassical approximation, leading to a universal power law for the
energy loss into radiation in the asymptotics.
Quantum
origin of the effect manifests here rather in vanishing energy of the final
soliton state and in the model-dependent tunneling effects, which
determine the collapse probability $P_{coll} \simeq \exp(-S/\h)$, where
the value of $S$ can be zero as well.

To avoid unnecessary
complications we'll consider  these effects within the generic
1+1-dimensional model of a nonlinear scalar field $\varphi(x,t)$,
described by the Lagrangean density $$ \lagr(\f)=\1/2 (\pd_\m \f)^2 -
U(\f),  \eqno (1)$$ which possesses a classical solitary wave solution
$$\f_{c}(x,t)=u\({x-vt \over \sqrt{1-v^2}}\). \eqno(2)$$
Soliton quantization with special emphasis on
relativistic properties is implemented by means of the exact covariant
center-of-mass coordinate [4-9],
which in 1+1-dimensions  is given by $$ q=
\1/2 \({1 \over H}L+L{1 \over H}\), \eqno(3)$$ where
$L$ is the generator of  Lorentz transformations and $H$ is the
total Hamiltonian of the system. The operators $H,L$ and the
generator of the spatial translations $P$ obey the Lorentz algebra $$
[L,H] =i\h P,\quad [L,P] =i\h H,
\quad [H,P]=0. \eqno(4) $$

The general analysis says, that the c.m.s.
coordinate enters the Heisenberg operators always in the combination
$x-q(t)$, while the spatial translation is induced by $q \to q+a$ [5,7-9].
  For our purposes it would be more convenient to introduce $q$ in a
dimensionless form by means of the substitution $$x \to \x(x,t),
\eqno(5)$$ where $$ \x(x,t)=H^{1/2}(x-q(t))H^{1/2}=Hx-Pt-L.
\eqno(6)$$
So in the vicinity of the soliton solution the Heisenberg field
$\f(x,t)$ is represented as
$$ \f(x,t)=f(\x(x,t))+\F(\xi(x,t),t).  \eqno(7)$$ In eq.(7) we
take care of that the operator-valued argument of the soliton field might
affect the soliton shape to be different from the classical one
and therefore use new notation $f(\x)$ for it instead of that in (2).
The meson
field $\F(\xi,t)$, considered as a function of c-number arguments, is
taken independent of the generator $P$ of spatial translations of the
algebra (4), while the commutation relations between $H,L$ and
 $\F(\xi,t)$ are determined in such a way, that enables to provide the
 covariance  of the whole field.  The general procedure of such type has
been considered in detail in refs. [6-9].  However, within the present
approach we'll deal mainly with the soliton field $f(\x(x,t))$, which is
a covariant operator by construction. Note also, that  $\xi(x,t)$ commutes
with the mass operator $M^2=H^2-P^2$, and so $M$ can be considered as a
c-number by dealing with operators of the type $f(\x(x,t))$.

It can be argued, that the substitution (5-7) represents the general form of
the Heisenberg field in the one-soliton sector. Indeed, it was shown in
[5], that the replacement (5) appears as a direct result of
summing up the pertinent terms in the perturbative solution of equations
of motion starting from the single classical soliton. It can be also shown
[7-9], that within the expansion in inverse powers of the soliton
mass this method is equivalent to the canonical quantization via
collective coordinate [11-13]. A quite different relativistic
framework for soliton quantization based on the BRST approach has been
proposed recently in refs.  [14,15].  However, the latter method
turns out to be not quite appropriate to study quantum effects
considered below, that are essentially non-analytic in $\h$ and so lie
beyond the quasiclassical perturbation expansion.

The main consequence of the substitution (5) is that in the resulting
effective theory all the differential operators of the initial theory are
replaced by finite differences of special form. Namely, the dynamical
equation, that determines the invariant soliton shape $f(\x)$,
reads [4] $$ f(\x +i
\h)+f(\x -i\h) -2f(\x)+{\h^2 \over \mu^2} V'(f(\x))=0, \eqno(8)$$ where
it is convenient to extract the dimensional (mass) parameter $m$ from the
interaction term and to introduce
the dimensionless soliton mass $$ U(\f) = m^2 V(\f), \quad \m=M/m. \eqno(9) $$
Then the eq.(8) is written completely in terms of
dimensionless variables. (Recall, that by means of a suitable redefinition
of variables [1,4], in 1+1-dimensions we can always treat $\h$ as the
dimensionless expansion parameter for the soliton sector.)

The mass of the soliton is determined  from the following equation
  $$ \m^2={\m^2 \over 2\h^2} \int \!\! d\x \ [
 f(\x+i\h)-f(\x)][f(\x-i\h)-f(\x)]+ \int \!\!  d\x \ V(f(\x)) \ ,
 \eqno(10)$$ where the status of $f(\x \pm i\h)$ is the same as in
 eq.(8). The detailed discussion of eqs.(8) and (10) is given in ref.
[16].

The main consequence of the finite-differences in the dynamical equations
is the appearance of what should be called quantum copies of classical
solutions [4].
Namely, seeking the solution of eq.(8) in the form of the exponential
(Dirichlet) expansion, we obtain the following series of
approximate solutions, valid for $\h$ sufficiently small, namely $$
f_N(\x)=u(\a_N\x), \eqno(11)$$ where $N$ is an integer, $$ \a_N={1 \over
\m_N} + {2 \p N \over \s \h}, \eqno(12) $$ $\m_N=\m[f_N] $ is the
corresponding dimensionless mass of the field, and $\s$ is equal to the
mass of the elementary excitation of the field (meson) divided by $m$.

For $N=0$ the singular part in $\a_N$ disappears and the corresponding
solution $f_0(\x)=u(\x/\m_0)$ survives the  limit $\h \to 0$.
So
$f_0(\x)$ should be considered indeed as the quantum descendant of the
classical solution, and so in what follows the case $N=0$ will be referred
to as the quasiclassical one.
The masses $\m_N$ of soliton copies are determined from
eq.(10) and for small $\h$ and $N \not= 0$ are given by
$$ \m_N=\sqrt {\m_0 \over
|\a_N|} \simeq \sqrt{ \h \over |N|} \( {\m_0\s \over 2
\p} \)^{1/2}, \eqno(13)$$
while $\m_0$ is the classical soliton mass
$$\m_0=\int \! dx \ u'^2(x). \eqno(14)$$
The most essential point here is, that   $\m_N$ turn out to be $O(\sqrt
{\h})$  in magnitude, hence sufficiently smaller, than the classical mass
$\m_0$, but larger, than the typical meson energy, that is of order
$O(\h)$.  Another peculiarity in the result (13) is that $\m_N$ decrease
for increasing $|N|$. These properties will be crucial for the soliton
collapse.

Now let us consider these and other properties of quantum copies
using as an illustrative example the spontaneously broken $\f^4$-model,
that is given by the selfinteraction potential $ V(\f)=\1/2
(1-\f^2)^2$, and yields the classical kink solution $ \f_c(x)=\tanh x$. The
nonclassical kinks are given by
the shape functions $$ f_N(\x)=\tanh \a_N\x, \eqno(15)$$ where
$\a_N=1/\m_N + \p N/\h.$ (Recall, that in the $\f^4$-model with such
potential the mass of the elementary meson is $2m$, hence $\s=2$.)

An important property of quantum copies is that for
all $N \not= 0$ they carry the same topological charge $Q$ as the
classical soliton. For
$\h/N \to 0 \quad f_N(\x)$ converge  to the limiting
shape $f_{\infty}(\x)$ in the form of the discontinuous step-like
classical solution of the same topology. In the case of an odd classical
solution of the $\f^4$-type, what includes the most important soliton
models, we can write without loss of generality $$ f_N(\xi) \to
f_{\infty}(\x)=\sign \x \ . \eqno(16) $$ According to eq.(13), the mass of
$f_{\infty}(\x)$ vanishes.  This result is essentially nontrivial, since
in the purely classical case the kinetic energy of the discontinuous
solution like $\sign x$ is proportional to $\int \!  dx \ \d^2(x)$ and
diverges.  The reason is, that the energy of quantum copies is determined
by the operator fields $f_N(\x(x,t))$, rather than by the c-number shape
functions $f_N(\x)$. In particular, for sufficiently small $\h$ (large
$N$) the nonclassical kinks (15) can be approximated as $$f_N(\xi) \simeq
\tanh \( {\pi N \over \h} \xi \), \eqno(17)$$ hence are $i\h$-periodic,
and so by substituting them into the differences $[f(\x \pm i\h)-f(\x)]$
of eq.(10) the latter vanish identically.  Thus the quantum origin of
soliton copies is crucial for such specific behaviour of their
masses.

On account of these properties of quantum copies
we should
expect, that any given soliton state will be unstable under transition
with $\D N >0$, while emitting a corresponding number of mesons.  Since the
topological charge $Q$ remains unchanged, it should be just the soliton
collapse, rather than the soliton decay, that proceeds with $\D Q \not=
0$.  Now let us present the general analysis of transition from the
quasiclassical soliton state into its Nth quantum copy.  It would be
natural to describe such process in terms of suitable dynamical variables,
that continuously transform the initial configuration $f_0(\x)=u(\a_0 \x)$
into the final one $f_N(\x)=u(\a_N \x) $.  The simplest path
without infinite potential barriers, that enables such transition, is
given by the (linear) superposition of initial and finite soliton states,
both taken in the rest frame, $$ f(\x,t)=c_0(t)f_0(\x)+c_N(t)f_N(\x) \ ,
\eqno(18)$$ where the finite energy of the intermediate configurations is
provided by the subsidiary condition, imposed on scaling parameters,
namely $$ c_0(t)+c_N(t)=1. \eqno(19)$$ So the path (18) reduces to a
one-parameter family of trajectories $$ f(\x,t)=f_0(\x)+c(t)\D f(\x) \ ,
\eqno(20)$$ where $$ \D f(\x)=f_N(\x)-f_0(\x) \eqno(21)$$ is a regular
function with (exponentially) decreasing asymptotics.  The collapsing
process is given by the continuous evolution of $c(t)$ subject to boundary
conditions $$ c(0)=0 \ , \ \ \hbox {$\dot c $}(t)=0 \ ; \quad c(\infty)=1
\ , \ \ \hbox {$ \dot c $}(\infty)=0. \eqno(22)$$

It
would be natural to assume, that the collapse is an adiabatic process
(what is confirmed by explicit calculation below).
Then the energy
loss  into radiation will be very low, hence $\F(\x,t)$ will be  small
compared to $f(\x,t)$.  Granted this, the meson field can be treated
perturbatively in the slowly varying c-number soliton background
$f(\x,t)$, while the back reaction of $\F(\x,t)$ on the
dynamics of $c(t)$ being neglected.  So in the first step we have to
consider the proper dynamics of $c(t)$ without radiation. For these
purposes we insert the collapsing soliton field (20) into the initial
field Hamiltonian and evaluate the energy (mass) of the configuration as
an explicit function of $c(t)$ and $ \hbox {$ \dot c $}(t)$. The result
looks like eq.(10) by adding the terms with $ \hbox {$ \dot c$}(t)$,
namely $$ \m^2= \1/2 \dot{c}^2(t)  \int \!  d\x \ \[\D f(\x) \]^2 + \1/2
{\pd \over \pd t} \left\{ \dot {c}(t)  \int \!  d\x  \ \[\(1-\cos \h
\pd_\x\) f(\x,t)\] \ \D f(\x) \right\} +$$ $$+ \int \!  d\x \ \ham (\x,t)\ ,
\eqno(23)$$ where $$ \ham (\x,t)= {\m^2 \over 2\h^2} \left\{ \[ \( 1-\cos
 \h \pd_\x \) f(\x,t) \]^2+ \[ \( \sin \h \pd_\x \) f(\x,t) \]^2 \right\}
+ V(f(\x,t)).  \eqno(24)$$ Note, that by integrating over $d\x$ in
eqs.(23) and (24) we can always take the final configuration as the
discontinuous $f_{\infty}(\x)$. It is
definitely correct for transition to $N \to \infty$, but for $\h$ small
such replacement would be valid actually for any transition from the
quasiclassical state into $N\not= 0$.

Calculation of the finite-difference kinetic term  $T$ in $\int
\! d\x \ \ham(\x,t)$ proceeds by means of the following relations [4],
valid for any Nth soliton copy  $$ \( \sin \h \pd_\x \)  f_N(\x) \simeq
{\h \over \m_N} u'(\a_N\x), \quad \(1-\cos \h \pd_\x \)  f_N(\x) \simeq
\1/2 {\h^2 \over \m_N^2} u''(\a_N\x) \ , \eqno(25)$$ where the prime
denotes the derivative with respect to the argument, and the residual in
both expressions (25) is $O(\h)$. Therefore, for $\h$ small the term $
\[\(1-\cos \h \pd_\x \) f_N \]^2$ can be certainly neglected while
compared to $ \[\( \sin \h \pd_\x \) f_N\]^2$. Then on account of the
eq.(10) we obtain for  the kinetic term of the eq.(24) the following
expression $$ \int \! d\x \  \[\( \sin \h \pd_\x \) f(\x,t)\]^2 =
\h^2(c_0^2+c_N^2)+2{\h^2 \over \m_0\m_N} \ c_0c_N \int \! d\x \
u'(\a_0\x)u'(\a_N\x). \eqno(26)$$
The last term in eq.(26)
turns out to be $O(\h^{5/2})$, hence in $\sqrt{\h}$ times less than the
first one, and so can be dropped to the lowest order. So the answer for
$T$ acquires a very simple form
$$T= {\m^2 \over 2} (c_0^2+c_N^2)={\m^2 \over 2} \[c^2+(1-c)^2\].
\eqno(27)$$ Applying the same procedure to the second term in the eq.(23),
we can estimate it as $O(\h^2)$, and so for an adiabatic process with
bounded derivatives of $c(t)$ this term can be neglected as well.  So
finally the resulting expression for the energy of the collapsing soliton
configuration (20) can be written in terms of $c(t)$ and $\hbox{$ \dot c
$}(t)$ as $$ \m^2= {\g \over 2} \dot {c}^2+{\m^2 \over 2}\[c^2+(1-c)^2\]
+W(c) \ , \eqno(28)$$ where $$ \g=\int \! d\x \ \[\D f(\x)\]^2, \quad
W(c)=\int \! d\x \ V\[f_0(\x)+c\D f(\x)\] \ . \eqno(29)$$ For the
$\f^4$-model the straightforward evaluation of integrals in eqs.(29) gives
$$ \g={8 \over 3} (2 \ln 2-1) \ , \quad  W(c)= {4 \over 3} (c-1)^2 \[ \(8
\ln 2 - {16 \over 3}\)c^2+{2 \over 3} c+ {2 \over 3} \] \ . \eqno(30)$$

Now let us consider more carefully
the most general features of the
behaviour of $c(t)$, that are irrespective of the specific properties of
the model. Firstly, by definition  $W(c=1)=0$, hence
the point $c=1$ corresponds to the absolute minimum of $W(c)$, while
$W(c=0)=\m_0^2/2$. So the point $c=1$ will always lye within the
classically allowed region for the motion of $c(t)$, whose velocity at
this point is $  \dot c^2 =\m^2/ \g$.

Further, with respect to the shape of $W(c)$ on the interval
$[0,1]$ one has to distinguish between two different situations. The first
one concerns the case, when both the points $c=0$ and $c=1$ belong to the
same classically permissible region of motion. Then $c(t)$ will swing
nonlinearly between the neighboring turning points, that are given by
$c_1=0$ and $c_2 >1$, while the point $c=1$ will be the intermediate point
of the trajectory.  Indeed such
 a situation takes place in the $\f^4$-model.

The second case corresponds to the situation, when $W(c)$ yields a finite
potential barrier between $c=0$ and $c=1$. Now in order to reach the final
point
$c=1$ while starting from $c=0$, we have firstly to penetrate the barrier,
what gives the quantum-mechanical tunneling amplitude $P_{tunn}$ for the
probability of the whole collapsing process. As soon as we penetrate
the barrier and so reach the classically permissible region, which
the point $c=1$ belongs to, there starts the process of nonlinear
oscillations between the points $0<c_1<1<c_2$, similar to described above.
Since the tunneling proceeds at the energy close to the initial classical
soliton mass $\m_0$, that is large enough to use the quasiclassical
picture, the tunneling amplitude can be estimated through the conventional
WKB-approximation $$ P_{tunn} \simeq e^{-S/\h},
\eqno(31)$$ where $S$ is the corresponding classical action for
the forbidden region.

The most strict predictions can be made for the asymptotic behaviour of
the collapse, when  $\m$ is sufficiently small already and so the
oscillations shrink to a small neighborhood of the minimum of
$W(c)$, i.e. to the point $c=1$. Then we can replace the second term in
the eq.(28) by $\m^2/2$, simplifying it up to
$$ {\m^2 \over 2}= {\g \over 2} \dot {c}^2+W(c). \eqno(32)$$
Now let us write $$ c(t)=1+ \tilde{c}(t) \ , \eqno(33)$$ where
$\hbox {$ \tilde c $}(t)$ is small, and represent the collapsing configuration
as
$$ f(\x,t)=f_{\infty}(\x)+\tilde{c}(t) \D f(\x).  \eqno(34)$$
Since $V(f_{\infty})=V'(f_{\infty})=0$, from the eq.(34) we observe at once,
 that the expansion of $W(c)$ in powers of $\tilde c$ is equivalent to the
expansion around the true vacuum value, what gives $$ \left. \m^2 \right|
_{\m \to 0}=\g \[ {\dot {\tilde c} \hbox{\hskip 0.1 em \vrule width 0mm
height 2.0 ex } }^2 + \s^2 \tilde {c}^{2} \] \ . \eqno(35)$$
As a result, the limiting oscillations will be harmonic with
the frequency $$ \n=\s \ .  \eqno(36)$$

Such peculiar behaviour of the asymptotical dynamics of the system gives
rise to one more general feature of the collapsing process. Let us assume,
that there exists a set of discrete modes (bound states) $\phi_n(x)$ in
the spectrum of meson excitations around the soliton.
Then it might seem, that the transition
from the initial to finite soliton states might occur through the
excitation of certain $\phi_n$. However, actually such a picture doesn't
take place, since the eigenfrequencies of the discrete spectrum lye always
lower, than the meson mass, namely $0<\w_n<\s$, and so due to the relation
(36) the resonance in the meson excitation will be just at the threshold
of the continuum.  Therefore, the excitation of the bound states will be
suppressed compared to the creation of wavepackets, which remove the
energy from the soliton.  Moreover, it can be argued [16], that the
mesonic modes around the final soliton state fall out of the dynamics at
all.

To estimate the energy loss into radiation we have to look at the dynamics
of the meson component $\F(\x,t)$ during the collapse.
We start with the obvious statement, that the energy of the full
Heisenberg field $\f(x,t)$, that includes both the collapsing soliton and
the meson field, certainly conserves. It means, that
if the starting point of the collapse is $c=0, \ \dot c=0, \ \m=\m_0$, the
total mass of the field remains equal to $\m_0$ throughout the process.
In turn, in the rest frame the wave operator $\dalam_{xt}$
acquires in terms of finite differences (after suitable redefinition of
the time $t \to mt$) the following form $$ \dalam_{xt} \to
\pd_t^2+2{\m_0^2 \over \h^2} \(\cos \h\pd_\x-1 \).  \eqno(37)$$ As a
result, the initial field equation, whence the dynamics of $\F(\x,t)$
should have been determined, can be written as $$ \[ \pd_t^2+2{\m_0^2
\over \h^2} \(\cos \h\pd_\x-1 \)\] \tilde \f (\x,t)+V'(\tilde \f(\x,t))=0
\ . \eqno(38)$$ In the eq.(38) $\tilde \f (\x,t)$ stands for the
Heisenberg field transformed to the $\x$-reference frame $$ \f(x,t)=\tilde
\f (\x(x,t),t), \eqno(39)$$ whereas during the collapse $$ \tilde \f
(\x,t)=f(\x,t)+\F(\x,t).\eqno(40)$$

The direct consequence of such initial conditions is that $f_0(\x)$
satisfies the eq.(38) with the required precision.  Therefore
the region  $|c|\ll 1$, where
$f(\x,t)$ is close to $f_0(\x)$, cannot significantly contribute to the
radiation per period of oscillations.  The dominant contribution
originates from the neighborhood of $c=1$, where $f(\x,t) \simeq
f_N(\x,t)$. During the initial stage of the process such a picture will
hold, because $f_N(\x)$ doesn't satisfy the eq.(38), whereas for $t$
large the oscillations of $c(t)$ will shrink to a small neighborhood of
$c=1$.   There remains, of course, an intermediate region, where the
oscillations won't be harmonic yet, while their amplitude will be large
enough for the nonlinear effects to be significant.  However,  in the
asymptotics the radiation will be indeed very small, while at the
beginning of the process the nonlinear terms cannot play any significant
role, since it takes some definite time for any nonlinearity  to develop
[17].  Therefore these effects can be essential only for a finite interval
of time, and so cannot seriously affect the main properties of the
process, while depending to a high degree on the specific features of the
model.  So in order to retain the generality we'll omit this nonlinear
region.

Now let us define more correctly the status of the meson excitations
during the collapse.  First of all, since the meson field must
compensate the c-number difference in soliton shapes $\D f(\x)$, to the lowest
order $\F(\x,t)$ should be treated as a c-number
coherent wave.  Further, any mesonic excitation should be referred to the
initial soliton state with $N=0$, what we explicitly take account of by
putting $$ \F(\x,t)=\tilde \F(\a_0 \x, t).  \eqno(41)$$ Making use of
relations (25), we can now easily verify, that for such a function the
finite-difference wave operator (37) can be replaced by $$ \[
\pd_t^2+2{\m_0^2 \over \h^2} \(\cos \h\pd_\x-1 \) \]\F(\x,t) \to
\dalam_{yt}\tilde \F(y,t), \eqno(42)$$ where $y=\a_0\x$. Actually, the
last relation is nothing else, but the manifestation of the quasiclassical
nature of the sector with $N=0$.

After these preliminary remarks we insert the decomposition (40) into the
eq.(38)  and linearize it with respect to $\F(\x,t)$ in the vicinity of
$f_N(\x)$, while supposing, that the main contribution to $\F(\x,t)$
appears from the neighborhood of $c=1$.
Then we get the simplified equation
$$ \[  \ \dalam_{yt}+V''(f_N)\ \]\tilde \F=j, \eqno(43)$$
where the radiation generating current $j$ is determined as the residual
of the eq.(38), applied to the collapsing soliton field in the vicinity of
$f_N(\x)$ $$ -j(\x,t)= \ddot{c}(t) \D f(\x)+ 2{\m_0^2 \over \h^2} \[
\(\cos \h\pd_\x-1 \)f_N(\x)\]+V'[f_N(\x)]. \eqno(44)$$ By means of the
relation (25) the difference term in $j(\x,t)$ can be represented as $$ 2
\ {\m_0^2 \over \h^2} \[\(\cos \h\pd_\x-1 \)f_N(\x)\] \simeq -{\m_0^2
\over \m_N^2} \ u''(\a_N\x). \eqno(45)$$
It is easy to verify, that for $\h/N \to 0$ we can write $$ {\m_0^2 \over
\m_N^2} \ u''(\a_N\x) \to {Q \m_0 \over \a_N} \ \d'(\x),  \eqno(46)$$
where $Q$ is the topological charge and so doesn't depend on $\h$. The factor
$\a_N^{-1}$ in the r.h.s.  of the eq.(46) survives the integration over
$d\x$ with the smooth function of the type (41), but vanishes for $\h/N
\to 0$.  So the difference term in the radiation current can be dropped.

Finally, we replace $f_N(\x)$ in the eqs.(43) and (44) by the limiting
$f_{\infty}(\x)$, for which
$ V'(f_{\infty})=0, \ \ V''(f_{\infty})=\s^2.$
So in the lowest approximation the eq. of motion for the meson field reads
simply
$$ \( \dalam_{yt}+\s^2\)\tilde \F(y,t)=-\hbox{$\ddot c $}(t) \D f(\x).
\eqno(47)$$
Now, to avoid the excess of precision we retain in $\D f(\x)$ only
the first term in the asymptotics for $ |\x| \to \infty$. Then for
the case of an odd classical soliton
we can represent $\D f(\x)$ explicitly as
$$ \D f(\x) \simeq  e^{-\s|y|} \sign y , \quad y=\a_0\x. \eqno(48)$$
Motivated by the same reasons, in the last step  we keep only the main
harmonics of $c(t)$ with the frequency $\n=\s$,  approximating the
oscillations by
$$ c(t)=1+A\cos \n t. \eqno(49)$$
Then the eq.(43) can be easily solved, whence for the radiation
energy $\D \E$ per the main period of oscillations $\D t=2 \p /\n$ we find
$$ \D \E=4 {\(A\n^2\)^2 \over \p} \int \! {dk \over k^2 (k^2+\s^2)} \
\sin^2 \[ {\D t \over 2} \sqrt {k^2+\s^2} \]
=3.926 \ { A^2 \n \over \p}. \eqno(50)$$

Thus, the radiation energy per period  can be
estimated as $\D\E =\C \times A^2$, where $A$ is the magnitude of
oscillations around $c=1$. From this
relation the asymptotic rate of the collapse can be easily deduced while
noticing, that for sufficiently large $t$ the remaining soliton mass is
determined through the eq.(35), what gives
$$ \m^2=\g \ \n^2A^2. \eqno(51)$$
Taking the average over the period, we can write down now the following
equation for the energy loss
$$ {d\m \over dt}= -\D \E=-\C \times \m^2 \ ,\eqno(52)$$
where
$$ \C={3.926 \over \p  \n \g}. \eqno(53)$$
Therefore the asymptotical law for the remaining soliton mass $\m(t)$ reads
$$ \m(t)={\C \over t} \ , \eqno(54)$$
while the amplitude of oscillations around $c=1$ and the radiation
power decrease as $1/t$ and $1/t^2$ correspondingly.

So the soliton collapse turns out to be indeed a very slow process with
the time scale given by the constant (53). It should be stressed,
however, that this constant plays here a quite different role compared to
the decay time $\t$ of the quantum-mechanical quasistationary state, that
decays according to the exponential law $\exp\(-t/\t\)$. In the latter
case $\t$ determines the time interval, during which practically all the
excitation energy should have been emitted, whereas the soliton collapse
reveals a slowly decreasing tail of radiation.

The asymptotic rate of the collapse doesn't  depend
explicitly on the difference parameter $\h$ (or $\h/c$, when the speed
of light $c$ is restored explicitly). The reason is, that the underlying
quantum origin of the effect manifests here in the vanishing energy of the
final soliton state, whatafter the whole process can be successfully
formulated within the quasiclassical picture. However, there exists another
possibility for $\h$ to enter the resulting dynamics, namely through the
collapse probability $P_{coll}$.
When there are no barriers, as for the $\f^4$-theory, the
whole collapse proceeds in the real time, thence $$ P_{coll}=1.
\eqno(55)$$ On the contrary, in presence of a barrier one has firstly to
penetrate  it, what gives rise of the tunneling amplitude $P_{tunn}$, and
so the collapse probability can be estimated as $$ P_{coll}=P_{tunn}
\simeq e^{-S/\h} \ , \eqno(56)$$ although in both
cases the asymptotical behaviour is governed by the same power law (54).
So in presence of tunneling the collapse acquires an explicit exponential
dependence on $\h$, what apparently demonstrates the non-perturbative
origin of the effect.  Remarkably enough, this time $\h$ comes merely
from the purely quantum-mechanical considerations, hence it has nothing to
do with finite differences and so with the speed of light. So it turns
out, that while emerging as an essentially relativistic effect, the
collapse actually survives the non-relativistic limit $c \to \infty$.

To conclude let us note, that such a collapse turns out to be an immanent
feature of the particle-like classical solutions in the relativistic QFT.
The reason is, that the energy density of such extended objects is
localized, therefore the covariant c.m.s. coordinate is a well-defined
quantity (see [18-20] and references therein).  Then the finite-difference
structure of the effective dynamics in covariant coordinates implies, that
upon quantization  the mass of the limiting step-like configuration
$f_{\infty}(\x)$ will vanish, rather than diverge.  In this way the
extended particle-like objects induce their collapse by themselves.

Actually, the
derived asymptotic law (54) should be valid in more
spatial dimensions as well, since the latter
maintain the difference structure of the effective dynamics in covariant
coordinates [10].
Moreover,  the structure of
corresponding difference operators coincides with those of the
quasipotential approach on the mass shell in  momentum space
[21], what once more emphasizes the
relativistic origin of the effect.  So we should expect, that the main
features of the effect will survive the transition to more spatial
dimensions. The application of the general framework developed in [10] to
this problem will be reported separately.

This work is supported in part by the Russian Foundation of Fundamental
Research, Grant No. 94-02-05491,
by the Sankt-Petersburg Centre of Fundamental Investigations,
and by the Scientific Program "Russian Universities".

\vskip 0.5 true cm

\large \rm

\end{document}